\documentclass[twocolumn,showpacs,aps,amsmath,amssymb,superscriptaddress]{revtex4}
\usepackage{graphicx}
\begin{document}
\title{Alignment of Rods and Partition of Integers}
\author{E.~Ben-Naim}
\affiliation{Theoretical Division and Center for Nonlinear Studies,
Los Alamos National Laboratory, Los Alamos, New Mexico 87545}
\author{P.~L.~Krapivsky}
\affiliation{Department of Physics and Center for Molecular
Cybernetics, Boston University, Boston, Massachusetts, 02215}
\begin{abstract}
  We study dynamical ordering of rods. In this process, rod alignment
  via pairwise interactions competes with diffusive wiggling.  Under
  strong diffusion, the system is disordered, but at weak diffusion,
  the system is ordered. We present an exact steady-state solution for
  the nonlinear and nonlocal kinetic theory of this process. We find
  the Fourier transform as a function of the order parameter, and show
  that Fourier modes decay exponentially with the wave number. We also
  obtain the order parameter in terms of the diffusion constant. This
  solution is obtained using iterated partitions of the integer
  numbers.
\end{abstract}
\maketitle

\section{Introduction}

Phase transitions from an isotropic, disordered state to a nematic,
ordered state are fundamental in equilibrium statistical physics
\cite{hes}. They occur in liquid crystals \cite{cl,dgp}, complex
fluids \cite{sas}, and are closely related to phase synchronization
\cite{yk,shs,abprs}. The standard approach for describing
order-disorder phase transitions is based on a Hamiltonian
description. However, it may not necessarily apply in nonequilibrium
settings such as granular systems \cite{bdve,bnk,at1}, where a
non-equilibrium approach, that is based on kinetic interaction rules,
may be required.

We study dynamical alignment of rods and show that the corresponding
kinetic theory framework is appropriate for describing the underlying
order-disorder phase transition.  This approach conveniently allows
characterization of statistical properties such as the distribution of
rod orientation.

We consider the mean-field theory for alignment of interacting polar
rods, introduced recently as a model \cite{at} for ordering of
microtubules by molecular motors in a planar geometry
\cite{umamk,nsml,lk,kjjps,lm,SS}. In this model, rods become aligned
by pairwise interactions, and they may also wiggle in a diffusive
fashion, the latter process being governed by a white noise. This
system reaches a steady-state where alignment by interactions is
balanced by diffusion. With strong diffusion, the system is
disordered, but with weak diffusion, the system becomes ordered.

Our main result is an exact steady-state solution for this kinetic
theory. The Fourier modes obey a closed set of coupled nonlinear
equations. By iterating these equations, it is possible to relate all
Fourier modes to the lowest mode, which in turn is equivalent to the
order parameter. Similarly, the order parameter itself is shown to
obey a closed equation and this allows one to express the distribution
of rod orientations in terms of the diffusion constant. While in
principle, the complete solution involves an infinite number of
Fourier modes, in practice, it is sufficient to consider only a
moderate number of modes because the Fourier modes decay exponentially
with the wave number.

The exact solution is ultimately related to iterated partitions of the
integers numbers. This feature is generic: it applies to general alignment
rules, as well as to arbitrary alignment rates. We also find that depending
on the alignment rate, the system may or may not undergo a phase transition.

This paper is organized as follows. In Sect.~II, we describe the rod
alignment model. In Sect.~III, we perform the Fourier transform of the
governing equation, which is then solved in Sect.~IV.  We discuss
nearly perfect alignment in Sect.~V, and briefly mention several
generalizations in Sect.~VI. We conclude in Section VII.

\section{The Rod Alignment Model}

In the rod alignment model, introduced by Aranson and Tsimring
\cite{at}, there is an infinite number of identical polar rods, each
with an orientation $-\pi \leq \theta\leq \pi$. The rods become
aligned via irreversible pairwise interactions. As a result of the
interaction between two rods with orientations $\theta_1$ and
$\theta_2$, both rods acquire the average orientation as follows
(see also figure 1):
\begin{equation}
\label{alignment}
(\theta_1,\theta_2)\to
\begin{cases}
\left(\frac{\theta_1+\theta_2}{2},\frac{\theta_1+\theta_2}{2}\right)&
|\theta_1-\theta_2|<\pi\\
\left(\frac{\theta_1+\theta_2+2\pi}{2},\frac{\theta_1+\theta_2+2\pi}{2}\right)&
|\theta_1-\theta_2|>\pi.
\end{cases}
\end{equation}
Here, and throughout this study, addition and subtraction are
implicitly taken modulo $2\pi$.

\begin{figure}[h]
\includegraphics*[width=0.17\textwidth]{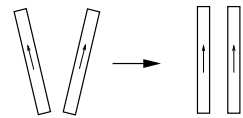}
\caption{Illustrating of the alignment process.}
\end{figure}

There is also randomness in the form of a white noise: each rod
wiggles in a diffusive fashion, and this process is characterized by
the diffusion constant $D$. Specifically, in addition to the alignment
process (\ref{alignment}), the orientation of a rod changes according
to $d\theta/dt=\eta$ with $\eta$ an uncorrelated white noise
$\langle\eta(t)\eta(t')\rangle=2D\delta(t-t')$.

Let $P(\theta,t)$ be the probability distribution function of rods
with orientation $\theta$ at time $t$.  It is normalized to one, 
$\int_{-\pi}^{\pi}d\theta\, P(\theta,t)=1$. This distribution function
satisfies the integro-differential master equation
\begin{equation}
\label{pt-eq}
\frac{\partial P}{\partial t}= D\frac{\partial^2
P}{\partial\theta^2} +\int_{-\pi}^{\pi}d\phi\,
P\left(\theta-\frac{\phi}{2}\right)
P\left(\theta+\frac{\phi}{2}\right)-P.
\end{equation}
The first term on the right-hand-side describes diffusion.  The
integral accounts for gain of rods with orientation $\theta$ as a
result of alignment of two rods with an orientation difference of
$\phi$, while the negative term accounts for loss due to
alignment. Without loss of generality, the alignment rate is set to
$1/2$ so that the loss rate equals one. This kinetic theory
generalizes the Maxwell model of inelastic collisions in a granular
gas as all rods interact with each other and additionally, the
alignment rate is independent of the relative orientation
\cite{bk1,bk2}.

\section{The Fourier Transform}

The governing master equation (\ref{pt-eq}) is nonlinear and nonlocal.
Its convolution structure suggests using the Fourier transform
\begin{equation}
\label{pk-def}
P_k=\big\langle  e^{-i k\theta}\big\rangle=\int_{-\pi}^{\pi}d\theta\,
e^{-i k\theta}\,P(\theta).
\end{equation}
The zeroth mode equals one, $P_0=1$, because of the normalization, and
also, $P_k=P_{-k}^*$.  The angular distribution can be expressed as a 
Fourier series
\begin{equation}
\label{pt-series}
P(\theta)=\frac{1}{2\pi}\sum_{k=-\infty}^{\infty} P_k\,e^{i k\theta}.
\end{equation}
Since the alignment process (\ref{alignment}) is invariant with
respect to an overall rotation $\theta\to \theta+\varphi$ with an
arbitrary phase $0\leq \varphi\leq 2\pi$, if $P(\theta)$ and $P_k$ are
solutions of (\ref{pt-eq}), then so are $P(\theta+\varphi)$ and
$P_ke^{i k\varphi}$, respectively.

The order parameter $R=\big|\langle
e^{i\theta}\rangle\big|=\left|P_{-1}\right|$, with the bounds $0\leq
R\leq 1$ probes the state of the system. A vanishing order parameter
indicates an isotropic, disordered state, while a positive order
parameter reflects a nematic, ordered state.

We focus on the steady state. Substituting the Fourier series
(\ref{pt-series}) into the master equation (\ref{pt-eq}), setting
$\partial P/\partial t\equiv 0$, and integrating over $\phi$, we find
that the Fourier transform satisfies a set of coupled nonlinear
equations \cite{at}
\begin{equation}
\label{pk-eq1}
\left(1+D k^2\right)P_k= \sum_{i+j=k} A_{i-j}\,P_i\,P_j,
\end{equation}
with $A_q=\frac{1}{2\pi}\int_{-\pi}^{\pi}d\phi\, e^{iq\phi/2}=
\frac{\sin \pi q/2}{\pi q/2}$. The coefficients $A_q$ are symmetric,
$A_q=A_{-q}$, and explicitly,
\begin{equation}
\label{aq}
A_q=
\begin{cases}
1&q=0;\\
0&|q|=2,4,\ldots;\\
(-1)^{\frac{q-1}{2}}\frac{2}{\pi|q|}&|q|=1,3,\ldots
\end{cases}
\end{equation}
In (\ref{pk-eq1}), when $k=0$, the sum contains only a single term,
$P_0=P_0^2$ and indeed, $P_0=1$.

Since $P_0=1$ is known, the identical terms $A_{-k} P_0 P_k$ and
$A_{k}P_k P_0$ in the steady-state equation (\ref{pk-eq1}) are linear
in $P_k$, and thus, we move them to the left-hand side.  Then
Eq.~(\ref{pk-eq1}) becomes
\begin{equation}
\label{pk-eq2}
\left(1+D k^2-2A_k\right)\, P_k=\sum_{\substack{i+j=k\\i\neq
0,\, j\neq 0}} A_{i-j}\,P_i\,P_j.
\end{equation}
Next, we absorb the prefactor on the left-hand side into the kernel on
the right-hand side by introducing the rescaled kernel
\begin{equation}
\label{gij}
G_{i,j}=\frac{A_{i-j}}{1+D(i+j)^2-2A_{i+j}}.
\end{equation}
With this transformation, we arrive at a compact, and
simpler-to-analyze, equation for the steady-state Fourier transform
\begin{equation}
\label{pk-eq}
P_k=\sum_{\substack{i+j=k\\i\neq 0,\, j\neq 0}}G_{i,j}P_i\,P_j.
\end{equation}
The kernel $G_{i,j}$ couples the $i$th and the $j$th Fourier modes. It
has the following properties: 
\begin{subequations}
\begin{align}
\label{gij-a}
G_{i,j}&=G_{j,i},\\
\label{gij-b}
G_{i,j}&=G_{-i,-j},\\
\label{gij-c}
G_{i,j}&=0,\qquad {\rm for}\qquad |i-j|=2,4,\ldots.
\end{align}
\end{subequations}
The governing equation (\ref{pk-eq}) is nonlinear and moreover, for
odd $k$, the sum contains an infinite number of terms. Despite this,
it is still possible to solve this equation analytically!

\section{Exact Solution}

First, we notice that Eq.~(\ref{pk-eq}) can be iterated once, leading
to a sum of products of three Fourier modes
\begin{equation}
\label{pk-eq3}
P_k=\sum_{\substack{i+j=k\\i\neq 0,\, j\neq 0}}
 \sum_{\substack{l+m=j\\l\neq 0,\, m\neq 0}}
G_{i,j}G_{l,m}\,P_i\,P_l\,P_m .
\end{equation}
Clearly, this procedure can be repeated any number of times, leading
to a sum over products of any given number of Fourier modes.  

Using this iteration procedure for all but the lowest Fourier mode, it
is possible to express $P_k$ as an explicit function of $P_{\pm 1}$
only. For instance, when $k=2$, Eq.~(\ref{pk-eq}) is simply
$P_2=G_{1,1}P_1^2$ and when $k=4$ one has $P_4=G_{2,2}P_2^2$. Thus,
the fourth Fourier mode is also expressed in terms of the lowest mode,
$P_4=G_{2,2}G_{1,1}^2 P_1^4$ \cite{note}. Odd modes, in contrast,
involve an infinite sum. For the third mode,
$P_3=2G_{1,2}P_1P_2+2G_{-1,4} P_{-1} P_4+\cdots$, and substituting the
second and the fourth moments above,
\begin{equation*}
P_3=2G_{1,2}G_{1,1}P_1^3+2G_{-1,4}G_{2,2}G_{1,1}^2P_1^4P_{-1}+\cdots.
\end{equation*}

Generally, the $k$th Fourier mode can be written as an infinite series
involving terms of the form
\begin{equation*}
P_1^{k+n}P_{-1}^n= P_1^k(P_1\,P_{-1})^n=P_1^k R^{2n}
\end{equation*}
with $n$ a positive integer. Recall that $P_1=R\,e^{i\varphi}$.
Without loss of generality, we can set the phase to zero,
$\varphi=0$. Then $P_1=R$ and the Fourier modes can be written
explicitly in terms of the order parameter $R$
\begin{equation}
\label{pk-sol}
P_k=R^k\,\sum_{n=0}^{\infty}\, p_{k,n}\,R^{2n}.
\end{equation}
Of course, $p_{0,n}=p_{\pm1,n}=\delta_{n,0}$. Since 
$p_{k,n}=p_{-k,n}$ it suffices to solve for $k>0$. 

The coefficients $p_{k,n}$ represent all iterated partitions of the integer
$k$ as follows
\begin{equation}
\label{partition}
k=\underbrace{1+1+\cdots+1+1}_{k+n}\underbrace{-1-\cdots-1}_n.
\end{equation}
The iterated partition involves a series of integer partitions
$k=i+j$, subject to the following restrictions: (i) $i\neq 0$ and
$j\neq 0$, (ii) $|k|\neq 1$, (iii) $G_{i,j}\neq 0$. Each such
partition contributes a factor $G_{ij}$ to $p_{k,n}$ (figure 2). For
example, the partition $2=(1+1)$ gives $p_{2,0}=G_{1,1}$ and the two
equivalent partitions $3=(1+2)=(1+(1+1))$ and $3=(2+1)=((1+1)+1)$ give
$p_{3,0}=2G_{1,2}G_{1,1}$. We list all non-vanishing terms up to the
fifth order in $R$
\begin{subequations}
\begin{align}
p_{2,0}&=G_{1,1},\\
p_{3,0}&=2G_{1,2}G_{1,1},\\
p_{3,1}&=2G_{-1,4}G_{2,2}G^2_{1,1},\\
p_{4,0}&=G_{2,2}G^2_{1,1},\\
p_{5,0}&=2G_{1,4}G_{2,2}G^2_{1,1}+4G_{2,3}G_{1,2}G^2_{1,1}.
\end{align}
\end{subequations}
Here, we used the symmetries (\ref{gij-a}) and (\ref{gij-b}) to
consolidate identical terms. We stress that the partitions used here
differ from traditional integer partitions in that they involve
negative integers and in that they are iterated \cite{ae}.

Substituting the series expansion (\ref{pk-sol}) into the steady-state
equation (\ref{pk-eq}) and equating the same powers of $R$, the
coefficients $p_{k,n}$ satisfy
\begin{equation}
\label{pkn}
p_{k,n}=\sum_{l+m=n}\sum_{\substack{i+j=k\\i\neq 0,\, j\neq 0}}
p_{i,l}\,p_{j,m}.
\end{equation}
Since the indexes $l$ and $m$ are positive, this is now a recursion
equation.  Starting with $p_{0,0}=p_{1,0}=1$, and utilizing the
symmetry $p_{k,n}=p_{-k,n}$, equation (\ref{pkn}) is solved
recursively. This provides a systematic method for obtaining the
coefficients $p_{k,n}$.

\begin{figure}[t]
\includegraphics*[width=0.18\textwidth]{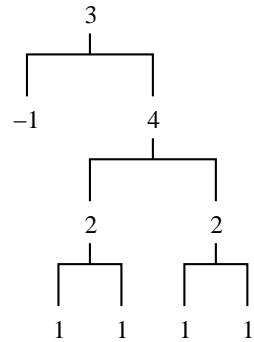}
\caption{The iterated integer partitions. Illustrated is the partition
corresponding to $p_{3,1}=2G_{-1,4}G_{2,2}G^2_{1,1}$. Each partition
$k=i+j$ with $i\neq 0$ and $j\neq 0$ generates a factor $G_{i,j}$. The
factor 2 accounts for the two equivalent partitions.}
\end{figure}

The series (\ref{pk-sol}) expresses all the Fourier modes in terms of
the order parameter $R$. It remains to obtain the order parameter $R$
as a function of the diffusion constant $D$. By Substituting the
Fourier solution (\ref{pk-sol}) into the governing equation
(\ref{pk-eq}) and setting $k=1$, we find that the order parameter
itself can be expressed as an infinite series
\begin{equation}
\label{r-sol}
R=\sum_{n=1}^\infty r_n R^{2n+1}.
\end{equation}
The coefficients $r_n$ are given by the very same recursion equation
(\ref{pkn})
\begin{equation}
\label{rn}
r_n=\sum_{l+m=n}\sum_{\substack{i+j=1\\i\neq 0,\, j\neq 0}}\,
p_{i,l}\,p_{j,m}.
\end{equation}
The coefficient $r_n$ is the counterpart of the coefficients
$p_{k,n}$ and it represents iterated partitions of the number $1$ as
in (\ref{partition}).  The partitions may not involve $0$'s.  Except
for the very {\it first} partition, the numbers $\pm 1$ may not be
repartitioned (Fig.~2). The first few coefficients are
\begin{subequations}
\begin{align}
\label{c1}
r_1&=2G_{-1,2}G_{1,1},\\
r_2&=4G_{-2,3}G_{1,2}G^2_{1,1},\\
r_3&=4(G_{-2,3}G_{-1,4}+G_{-3,4}G_{1,2})G_{2,2}G_{1,1}^3.
\end{align}
\end{subequations}

Equation (\ref{r-sol}) is a closed equation for the order parameter. Once it
is solved, all Fourier modes follow from (\ref{pk-sol}).  This completes the
solution for $R$ and $P_k$, and therefore for the angle distribution
$P(\theta)$.

In practice, one can calculate the order parameter $R$ as a root of a
polynomial of degree $N$ by truncating (\ref{r-sol}). For $N=3$,
substituting (\ref{aq}) and (\ref{gij}) into (\ref{c1}) yields
$r_1=-\frac{4}{3\pi}(1+D-4/\pi)^{-1}(1+4D)^{-1}$, and using
(\ref{r-sol}) gives the cubic equation for the order parameter
\begin{equation}
\label{cubic}
R=\frac{4}{3\pi}\frac{1}{D_c-D}\frac{1}{1+4D}R^3,
\end{equation}
with the critical diffusion constant
\begin{equation}
\label{dc}
D_c=\frac{4}{\pi}-1.
\end{equation}
For $D\geq D_c$, there is only the trivial solution $R=0$,
corresponding to an isotropic state where the rods are randomly
aligned: $P_k=\delta_{k,0}$ and $P(\theta)=\frac{1}{2\pi}$.  Below the
critical point, $D<D_c$, there is also the nontrivial solution
(Fig.~3)
\begin{equation}
\label{op}
R=\sqrt{\frac{3\pi}{4}(1+4D)(D_c-D)}.
\end{equation}
This corresponds to a nematic phase in which the rods are partially
aligned.  Near the transition point, this alignment is weak, but it
becomes stronger and stronger as $D$ decreases. The result of
Eq.~(\ref{op}) is approximate---only the cubic term in (\ref{r-sol})
has been maintained---yet sufficiently close to the transition point,
the corrections to the cubic equation (\ref{cubic}) are negligible and
$R\simeq C(D_c-D)^{\beta}$ with the prefactor
$C=\sqrt{(3/4)(16-3\pi)}$ and the mean-field exponent $\beta=1/2$
\cite{hes,yk,at} is asymptotically exact. As shown in the Appendix,
the uniform state is stable for $D>D_c$ but unstable for $D<D_c$.

\begin{figure}[t]
\includegraphics*[width=0.4\textwidth]{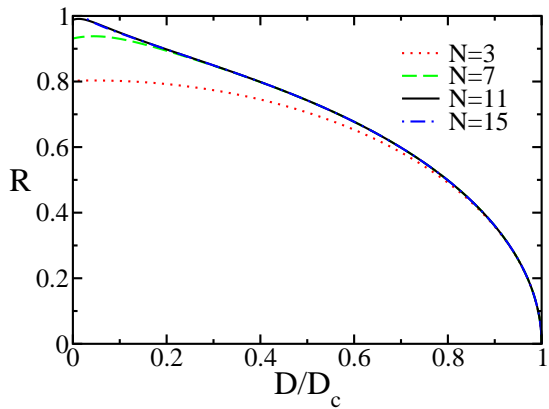}
\caption{The order parameter versus the diffusion coefficient. The
order parameter was obtained by solving polynomials of increasing
order.}
\end{figure}

Well below the critical point, we compute the coefficients
$p_{k,n}$ and $r_n$ to higher order from (\ref{pkn}) and
(\ref{rn}). The order parameter is then obtained by numerically
solving (\ref{r-sol}), truncated at the corresponding order. Since
the Fourier modes decay exponentially with the wave number
\begin{equation}
P_k\sim R^k,
\end{equation}
and since the order parameter obeys $0\leq R\leq 1$, a moderate number of
Fourier modes is sufficient to accurately compute $P(\theta)$. The order
parameter rapidly converges with $N$. For instance, $N=11$ already provided
an accurate value for $R$ (see Fig.~3). We note that at this order, it is
still possible to calculate the necessary partitions manually.

Once the order parameter is known, the Fourier modes are obtained from
(\ref{pk-sol}). The steady-state distribution (\ref{pt-series})
becomes
\begin{eqnarray}
\label{pt}
P(\theta)&=&\frac{1}{2\pi}+\frac{1}{\pi}\,R\cos\theta
+\frac{1}{\pi}G_{1,1}R^2\cos\,(2\theta)\\
&+&\frac{2}{\pi}G_{1,2}G_{1,1}R^3\cos\,(3\theta)
+O(R^4).\nonumber
\end{eqnarray}
In the vicinity of the transition point, the lowest mode dominates.
As the diffusion coefficient decreases, the angle distribution becomes
sharply peaked around $\theta=0$, reflecting that the rods are
strongly aligned (Fig.~4). We now analyze this nearly perfect
alignment in more detail.

\begin{figure}[t]
\includegraphics*[width=0.4\textwidth]{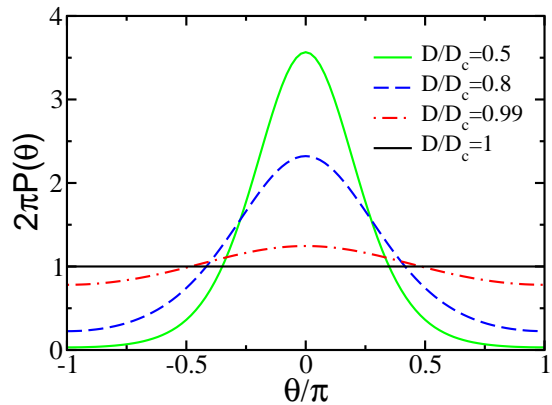}
\caption{The angular distribution for various values of $D$. The angle
distribution was obtained from the first 15 Fourier modes.}
\end{figure}

\section{Nearly perfect alignment}

For very small diffusion constants, the order parameter approaches
unity and therefore, the series solution (\ref{pk-sol}) is no longer
useful. Indeed, very close to $D=0$, the convergence is slow
(Fig.~3). Nevertheless, it is still possible to treat the problem by
performing the scaling transformation
\begin{equation}
\label{fx-def}
P(\theta)=\frac{1}{\sqrt{D}}f\left(\frac{\theta}{\sqrt{D}}\right).
\end{equation}
The scaling distribution remains normalized, \hbox{$\int dx f(x)=1$.}
At the steady-state, it obeys the nonlinear integro-differential
equation
\begin{equation}
\label{fx-eq}
f(x)=\frac{d^2f(x)}{d x^2}
+\int_{-\infty}^{\infty}dy\, f\left(x-\frac{y}{2}\right)
f\left(x+\frac{y}{2}\right).
\end{equation}
This equation was obtained by substituting (\ref{fx-def}) into the
master equation (\ref{pt-eq}), setting the time derivative to zero,
and replacing the integration limits $\pm\pi/\sqrt{D}$ with
$\infty$. In the limit $D\to 0$, the periodic nature becomes
irrelevant, and the problem reduces to the randomly forced inelastic
Maxwell model, for which several properties including the moments and
the Fourier transform can be obtained analytically \cite{bk1,bk2}.

Consider for example the normalized moments \hbox{$M_n=\langle
\theta^n\rangle/n!$}. They can be expressed in terms of the moments of
the scaling distribution $M_n=D^{n/2}m_n$ with $m_n=\int dx\, x^n
f(x)/n!$.  Substituting this definition into the master equation
(\ref{fx-eq}) and performing the integration, the moments $m_n$
satisfy the closed set of equations $m_n=m_{n-2}+2^{-n}\sum_{l=0}^n
m_l\,m_{n-l}$, from which the moments are found recursively, $m_0=1$,
$m_2=2$, and $m_4=\frac{18}{7}$ (the odd moments vanish). The order
parameter can be expressed in terms of the moments,
\hbox{$R=\big|\langle e^{i\theta}\rangle\big|=1-M_2+M_4+\cdots$} and
therefore, it is given by a Taylor series in powers of the diffusion
constant
\begin{equation}
R=1-2D+\frac{18}{7}D^2+\cdots.
\end{equation}
This expression is useful only for very small diffusion constants,
i.e., $D\to 0$.

Since the Fourier transform \hbox{$F(q)=\int_{-\infty}^{\infty} dx\,
 e^{iqx} f(x)$} is the generating function of the moments, it also
 satisfies a closed equation,
 $\left(1+q^2\right)F(q)=F^2\left(q/2\right)$.  Solving this equation
 recursively, the Fourier transform equals the infinite product
 \cite{bk1}
\begin{equation}
\label{fq-sol}
F(q)=\prod_{n=0}^\infty\left[1+\frac{q^2}{4^n}\right]^{-2^n}.
\end{equation}
The poles closest to the origin govern the large-$x$ asymptotics. The
simple poles located at $q=\pm i$ imply an exponential decay
\cite{bk2,se,adl}
\begin{equation}
\label{Pi-asymp}
f(x)\simeq Ce^{-|x|}
\end{equation}
as $|x|\to\infty$. The residues to these poles yield the prefactor:
\hbox{$C=\frac{1}{2}\,\exp\left[\sum_{n=1}^\infty
n^{-1}(2^{2n-1}-1)^{-1}\right]=1.47919$.}

\section{General Alignment Rates}

The above solution can be generalized in a number of ways. For
instance, it is straightforward to  generalize it to imperfect
alignment processes, where the orientation difference is reduced by a
fixed factor (analogous to an imperfect inelastic collision), in each
alignment event.

Moreover, it is also possible to analyze situations in which the
alignment process (\ref{alignment}) occurs with arbitrary alignment
rate $K\big(\theta_1-\theta_2)$. We assume $K(\phi)=K(-\phi)$ and
without loss of generality, impose the normalization,
$\frac{1}{2\pi}\int_{-\pi}^{\pi}d\phi\, K(\phi)=1$. The master
equation becomes
\begin{eqnarray}
0&=&D\frac{d^2 P}{d\theta^2}+
\int_{-\pi}^{\pi}d\phi\, K(\phi)P\left(\theta-\frac{\phi}{2}\right)
P\left(\theta+\frac{\phi}{2}\right)\nonumber\\
&-&
P(\theta)\int_{-\pi}^{\pi}d\phi\, K(\phi)P(\theta+\phi).
\end{eqnarray}
The Fourier modes satisfy a generalization of (\ref{pk-eq1})
\begin{equation}
\label{pk-eq1-gen}
Dk^2P_k=\frac{1}{2}\sum_{i+j=k} (A_{i-j}+A_{j-i}-A_{2i}-A_{2j})\,P_iP_j.
\end{equation}
The constant $A_q$ is now $A_q=\frac{1}{2\pi}\int_{-\pi}^{\pi} d\phi\,
e^{iq\phi/2} K(\phi)$.  Again $A_0=1$ and $A_q=A_{-q}$, but it is not
necessarily true that these constants vanish at even indexes as was
the case for uniform alignment rates. By excluding the zero modes
$i=0$ and $j=0$ from the summation in (\ref{pk-eq1-gen}), and by
following the same steps, we recover the governing steady-state
Eq.~(\ref{pk-eq}). The generalized coupling constants (\ref{gij}) are
\begin{equation}
\label{gij-gen}
G_{i,j}=\frac{1}{2}\frac{A_{i-j}+A_{j-i}-A_{2i}-A_{2j}}{1+D(i+j)^2-2A_{i+j}}.
\end{equation}
These coupling constants are manifestly symmetric $G_{i,j}=G_{j,i}$.
We conclude that the series solutions (\ref{pk-sol}) and (\ref{r-sol})
hold for arbitrary alignment rates.

By repeating the steps leading to (\ref{dc}), we find that the
critical diffusivity generally equals $D_c=2A_1-1$. The condition for
having a transition is therefore \hbox{$A_1>1/2$}.  For the constant
interaction rate \hbox{$A_1=\frac{2}{\pi}$}, and then
\hbox{$D_c=\frac{4}{\pi}-1$}. For the hard-sphere rate
$K(\phi)=\frac{2}{\pi}|\phi|$, then
\hbox{$A_1=\frac{4}{\pi^2}(2-\pi)$}, but since $A_1<1/2$, the system
is always disordered. Therefore, depending on the alignment rate,
there may or may not be an ordered phase.

We note that in the kinetic theory of gases, analytic solutions are
feasible only for the Maxwell model, where the collision rate is
velocity independent \cite{max,classical,e}, while the general
Boltzmann equation is not analytically tractable.  Remarkably, the
analytic solution presented above does not require a constant rate and
for example, the hard-sphere like collision rate $K(\phi)=C|\phi|$
{\it can} be solved analytically. The discrete nature of the Fourier
spectrum makes this possible.

\section{Conclusions}

In conclusion, we studied kinetic theory for alignment of rods. At the
steady-state, alignment by pairwise interactions is balanced by the
diffusive motion. At large diffusivities, the system is disordered and
at low diffusivities it becomes ordered.

The Fourier modes obey a closed equation that can be solved by
selective iterations. This allows one to express all Fourier modes in
terms of the lowest mode, or equivalently, the order
parameter. Similarly, one can obtain a closed equation for the order
parameter itself, and then the solution becomes explicit. Since the
Fourier modes decay exponentially with the wave number, a moderate
number of terms is sufficient to accurately obtain the full angular
distribution.

This kinetic approach fundamentally differs from the traditional
Hamiltonian approach for describing phase transitions from ordered to
disordered states. Yet, the characteristics of the phase transition,
including in particular, the critical exponent, are identical to the
mean-field theory. The kinetic theory has the virtue that it directly
yields distribution functions.

The exact solution, using integer partitions, is very general. It
applies to generic ``inelastic'' alignment rules and more importantly,
to general alignment rates. The periodic symmetry of the problem is
ultimately responsible for this because it involves a discrete Fourier
spectrum. It may be useful to generalize this approach to alignment of
rods in three dimensions.

The nonlocal and nonlinear governing equation, in its simplest form
(\ref{pk-eq}), appears in other physical processes including
aggregation \cite{fran}, and it may be interesting to utilize the
integer partition solution method in other contexts. There are other
natural extensions of this work including investigation of relaxation
toward the steady-state, studies of spatial correlations in
low-dimensional systems, and restricted range interactions where
multiple alignment directions may arise \cite{bkr,plr}.

Experimentally, it may be possible to realize this order-disorder
transition by vibrating macroscopic granular chains \cite{bdve} or
granular rods \cite{bnk,at1} where vibration causes diffusion and
inelastic collisions result in alignment.

\acknowledgments 
We thank Lev Tsimring for useful discussions.  We acknowledge US DOE
grant W-7405-ENG-36 and NSF grant CHE-0532969 for support of this
work.

\appendix
\section{Stability of the uniform state}  

The uniform state $P(\theta)=\frac{1}{2\pi}$ is always a steady-state
of the master equation (\ref{pt-eq}).  To check when this state is
stable, we write $P(\theta,t)=\frac{1}{2\pi}+p(\theta,t)$. To linear
order, the perturbation satisfies
\begin{equation}
\label{pert-eq} \frac{\partial p}{\partial t}=D\frac{\partial^2
p}{\partial\theta^2} +\int_{-\pi}^{\pi}\!\!d\phi\,
\frac{p(\theta-\phi/2)+ p(\theta+\phi/2)}{2\pi}-p.
\end{equation}
Let us take a periodic perturbation with wave number $k$ and growth
rate $\lambda$, that is, $p(\theta,t)\propto e^{ik\theta+\lambda
t}$. Substituting this form into Eq.~(\ref{pert-eq}) gives the
growth rate
\begin{equation}
\label{growth-rate} \lambda_k=2A_k-1-Dk^2.
\end{equation}
The growth rate is positive only for the lowest mode, $k=1$, and
hence, stability is governed by the smallest wave number $k=1$ for
which $\lambda_1=D_c-D$. Indeed, the uniform state
is unstable below the critical diffusion constant (\ref{dc}).

\end{document}